\documentclass[pra,onecolumn]{revtex4}%
\usepackage{amsfonts}
\usepackage{amsmath}
\usepackage{amssymb}
\usepackage{graphicx}%
\providecommand{\U}[1]{\protect\rule{.1in}{.1in}}

\begin{document}
\title{Quantum Nondemolition Measurement and Heralded Preparation of Fock States with
Electromagnetically Induced Transparency in an Optical Cavity }
\author{G. W. Lin$^{1}$}
\email{gwlin@ecust.edu.cn}
\author{Y. P. Niu$^{1}$}
\author{T. Huang$^{1}$}
\author{X. M. Lin$^{2}$}
\author{Z. Y. Wang$^{3}$}
\email{wangzy@sari.ac.cn}
\author{S. Q. Gong$^{1}$}
\email{sqgong@ecust.edu.cn}
\affiliation{$^{1}$Department of Physics, East China University of Science and Technology,
Shanghai 200237, China}
\affiliation{$^{2}$School of Physics and Optoelectronics Technology, Fujian Normal
University, Fuzhou 350007, China}
\affiliation{$^{3}$Shanghai~Advanced~Research~Institute,~Chinese~Academy~of~Sciences, Shanghai,~201210,~China~}

\begin{abstract}
We propose a technique for quantum nondemolition (QND) measurement and
heralded preparation of Fock states by dynamics of electromagnetically induced
transparency (EIT). An atomic ensemble trapped in an optical cavity is driven
by two external continuous-wave classical fields to form EIT in steady state.
As soon as a weak coherent field is injected into the cavity, the EIT system
departs from steady state, falls into transient state dynamics by the
dispersive coupling between cavity injected photons and atoms. Because the
imaginary part of time-dependent linear susceptibility $Im[X(t)]$ of the
atomic medium explicitly depends on the number $n$ of photons during the
process of transient state dynamics, the measurement on the change of
transmission of the probe field can be used for QND measurement of small
photon number, and thus create the photon Fock states in particular
single-photon state in a heralded way.

\end{abstract}

\maketitle

\emph{Introduction.}---Quantum nondemolition (QND) measurement, which is
designed to avoid the back action produced by a measurement \cite{Braginsky},
is of great importance in quantum information processing. For example, it
provides the possible method to resolve small photon number states without
changing them. Several QND measurement schemes have been proposed for
monitoring the small photon number $n$
\cite{Imoto,Holland,Brune,Imamoglu,Shimizu,Schuster}. In particular, Brune et
at. \cite{Brune} show that measurement of small photon number is possible
through Rydberg atom phase-sensitive detection. Recently, this method
\cite{Brune} has been experimentally demonstrated to be used for progressive
field-state collapse \cite{Guerlin}, quantum jumps of light recording the
birth and death of a photon \cite{Gleyzes}, and reconstruction of
non-classical cavity field \cite{glise}. In Ref
\cite{Brune,Guerlin,Gleyzes,glise}, the photons are stored in a mode of a high
quality microwave cavity and the QND measurement is made by the Ramsey
interference of nonresonant Rydberg atoms that are sent through the cavity one
by one. However, in the optical regime, QND measurement of photon number based
on cavity quantum electrodynamics (CQED) is still a challenge.

On the other hand, Fock states exhibit no intensity fluctuations and a
complete phase indetermination, which triggers off the research on the
preparation and nonclassical nature of photon Fock states
\cite{Hofheinz,Guerlin,Gleyzes,glise}. Single-photon Fock states in the
optical regime, as the excellent information carrier, are desirable for
applications in quantum information, e.g. unconditionally secure communication
in quantum cryptography \cite{Gisin}. Many theories and experiments have been
devoted to single-photon sources by accurately controlling a single emitter,
such as individual atom \cite{1}, molecule \cite{2}, nitrogen vacancy center
\cite{3}, or quantum dot \cite{4}. In reality, the weak coherent state pulses,
obtained by strongly attenuating a laser beam, are usually used for quantum
key distribution (QKD). Those weak coherent state pulses have a Poissonian
photon number statistics, and they present a large vacuum component, as well
as non-negligible multiphoton contributions, which could make long-distance
QKD suffer from a severe security loophole \cite{Brassard}.

In this letter, we propose a technique for QND measurement and heralded
preparation of photon Fock states in particular single-photon state, based on
dynamics of electromagnetically induced transparency (EIT) in an optical
cavity. EIT medium trapped inside a cavity has been explored for photon
blockade \cite{Imamoglu1}, vacuum-induced transparency \cite{Suzuki},
photon-number selective group delay \cite{Nikoghosyan}, NOON-state generation
\cite{Nikoghosyan11}, single atoms EIT \cite{cke}, and so on
\cite{Wu,Naeini,Albert}. In our scheme, an atomic ensemble trapped in an
optical cavity is driven by two external continuous-wave classical fields to
form EIT in steady state. As soon as a weak coherent field enters the cavity,
the EIT system departs from steady state, falls into transient state dynamics
by the dispersive coupling between cavity injected photons and atoms. Because
the imaginary part of time-dependent linear susceptibility $Im[X(t)]$ of the
atomic medium explicitly depends on the number $n$ of photons during the
process of transient state dynamics, the change of transmission of the probe
field can be used for QND measurement of the small photon number and heralded
preparation of photon Fock states. The scheme proposed here has the following
significant advantages: first, QND measurement in the optical regime is
achieved through measurement of the change of transmission of the probe field
in the dynamics of EIT, and the calculation shows QND measurement works in an
cavity without strict strong coupling; second, the preparation of photon Fock
states in particular single-photon state is probabilistic but in a heralded way.

\emph{Steady-state EIT in }$\Lambda$\emph{-type configuration.---}We first
review the three-level EIT in steady state \cite{Harris,Fleischhauer}. As
illustrated in Fig. 1, two continuous-wave classical fields, a probe field and
a coupling field with central angular frequency $\omega_{p}$ and $\omega_{c}$,
respectively couple two lower metastable states $\left\vert a\right\rangle $
and $\left\vert b\right\rangle $ to upper level $\left\vert e\right\rangle $,
and thus they form a standard $\Lambda$-type EIT configuration, in which the
coherent processes are described by interaction Hamiltonian (within the
rotating wave approximation)\begin{figure}[ptb]
\includegraphics[width=3.8in]{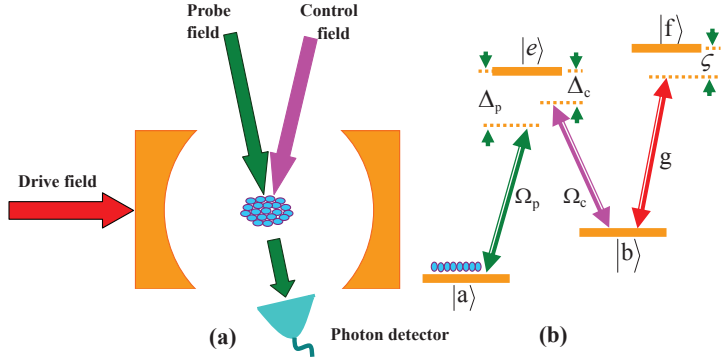}\newline\caption{(Color online) (a)
Schematic setup to QND measurement and the preparation of photon Fock states
based on dynamics of EIT. (b) The relevant atomic level structure and
transitions.}%
\label{1}%
\end{figure}%
\begin{equation}
H_{1}=-(\Omega_{p}\sigma_{ea}+\Omega_{c}\sigma_{eb}+H.c.)+\Delta_{p}%
\sigma_{ee}+\delta\sigma_{bb},
\end{equation}
here, $\Omega_{p}$ ($\Omega_{c}$) denotes the Rabi frequency associated with
the probe field (the coupling field), $\sigma_{\xi\eta}=\left\vert
\xi\right\rangle \left\langle \eta\right\vert $ ($\xi,\eta=a,b,e,f$) is the
atomic projection operator, $\Delta_{p}=\omega_{e}-\omega_{a}-\omega_{p}$,
$\Delta_{c}=\omega_{e}-\omega_{b}-\omega_{c}$ ($\hbar\omega_{\xi}$ is the
energy of the level $\left\vert \xi\right\rangle $), and $\delta=\Delta
_{p}-\Delta_{c}$. Under the two-photon resonant condition $\delta=0$, and when
the atomic ensemble has been driven into steady state, both real and imaginary
parts of the linear susceptibility vanish in the ideal case, which means that
the absorption and refraction of the probe field at the resonant frequency are
eliminated. That leads to the transparency in otherwise absorbed medium
\cite{Harris,Fleischhauer}.

\emph{QND measurement with dynamics of EIT.---} We consider such ensemble of
atoms are trapped inside an optical cavity as shown in Fig. 1(a), the atoms
couple to the quantized cavity field through the dipole coupling between level
$\left\vert f\right\rangle $ and $\left\vert b\right\rangle $. Thus the
interaction Hamiltonian is given by $H_{i}=\varsigma a^{\dagger}%
a+(g\sigma_{fb}a+H.c.)$, here $a(a^{\dagger})$ is the annihilation (creation)
operator of the cavity mode, $g$ is the coupling constant between the cavity
field and the atomic transition $\left\vert b\right\rangle \leftrightarrow$
$\left\vert f\right\rangle $, and $\varsigma$ denotes the detuning of the
cavity field from atomic transition $\left\vert b\right\rangle \leftrightarrow
$ $\left\vert f\right\rangle $. Suppose this coupling is in the case of large
detuning, where the upper atomic level $\left\vert f\right\rangle $ can be
adiabatically eliminated, the effective Hamiltonian is then given by
$H_{2}=Ga^{\dagger}a\sigma_{bb}$, with the dispersive coupling strength
$G=g^{2}/\varsigma$. We assume that under the two-photon resonant condition
$\delta=0$, the EIT system trapped in the cavity has been driven into steady
state. Then there is a photon Fock state $\left\vert n\right\rangle $ in the
cavity, the dynamics of the atom-photon density operator $\rho_{a-p}(t)$ is
governed by the master equation

\begin{figure}[ptb]
\includegraphics[width=6.0in]{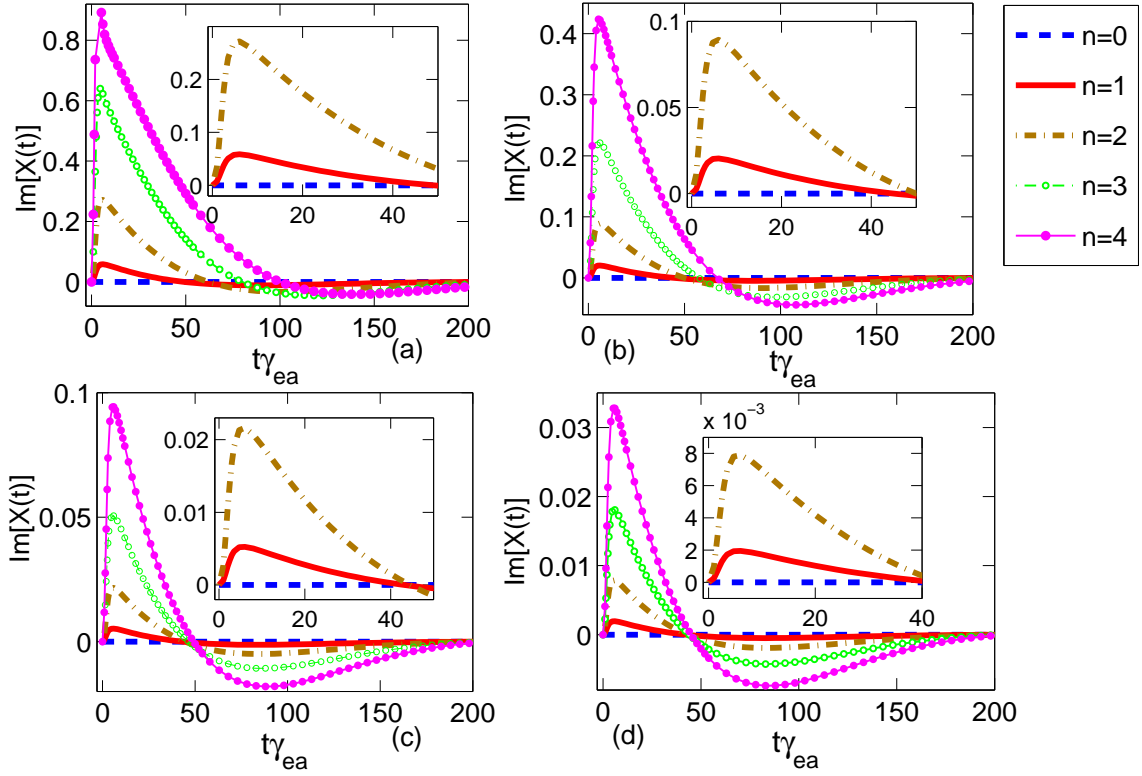}\newline\caption{(Color online) The
imaginary part of time-dependent linear susceptibility $Im[X(t)]$ versus the
time $t$ in units of $1/\gamma_{ea}$, with different Fock state $\left\vert
n\right\rangle $: $n=\{0,1,2,3,4\}$, when the dispersive coupling strength
$G=$ (a) $0.08\gamma_{ea}$, (b) $0.03\gamma_{ea}$, (c) $0.008\gamma_{ea}$, (d)
$0.003\gamma_{ea}$. Other common parameters are $\gamma_{eb}=\gamma_{ea}$,
$\gamma_{deph}=0.1\gamma_{ea}$, $\kappa=0.3\gamma_{ea}$, $\Delta_{p}%
=\Delta_{c}=\Delta=-\gamma_{ea}$, $N\left\vert \mu_{ae}\right\vert ^{2}%
/\gamma_{ea}\epsilon_{0}\hbar=1$, $\Omega_{p}=0.02\gamma_{ea}$, and
$\Omega_{c}=0.2\gamma_{ea}$.}%
\label{2}%
\end{figure}

\begin{figure}[ptb]
\includegraphics[width=5.0in]{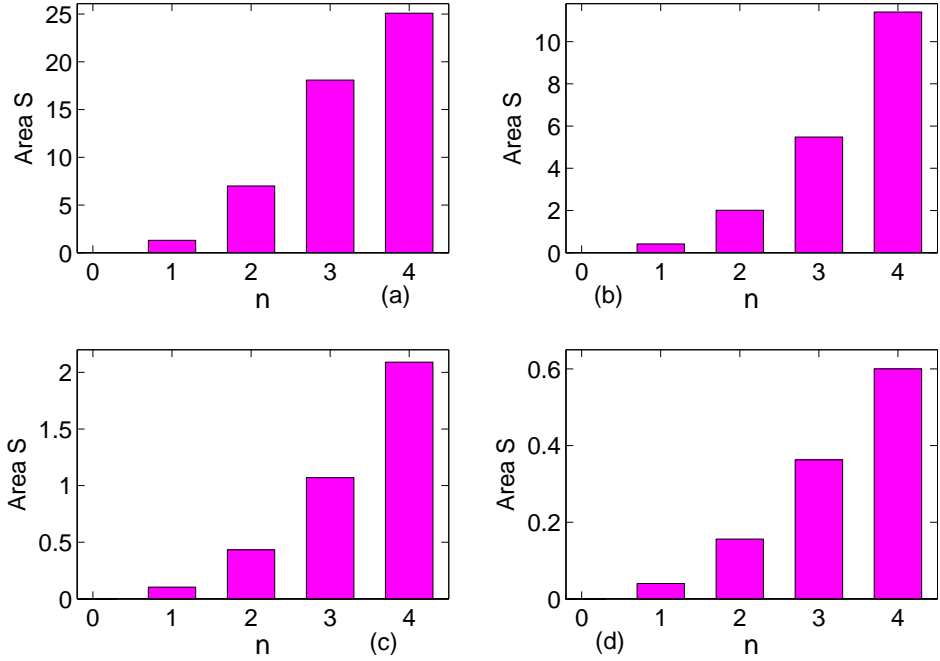}\newline\caption{(Color online) The
area $S={\displaystyle\int\nolimits_{0}^{T}}Im[X(t)]dt$ ($T=50/\gamma_{ea}$)
with the different dispersive coupling strengh $G=$ (a) $0.08\gamma_{ea}$, (b)
$0.03\gamma_{ea}$, (c) $0.008\gamma_{ea}$, (d) $0.003\gamma_{ea}$. Other
parameters are the same as in Fig. 2.}%
\label{3}%
\end{figure}

\begin{figure}[ptb]
\includegraphics[width=5.0in]{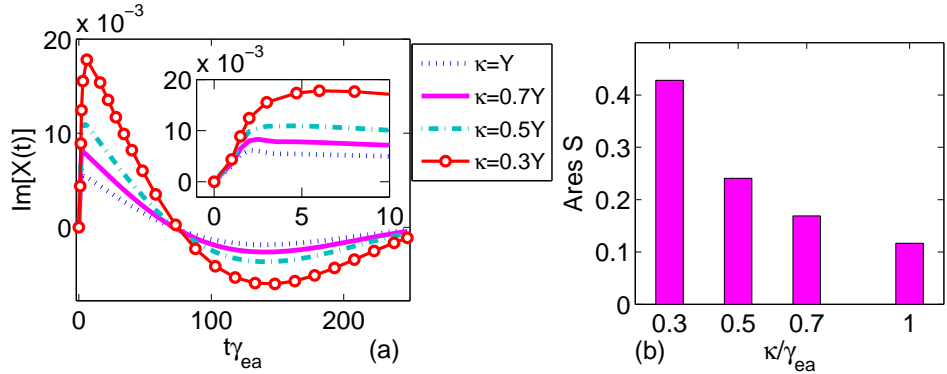}\newline\caption{(Color online)(a) The
imaginary part of time-dependent linear susceptibility $Im[X(t)]$ for the
single-photon Fock state $\left\vert 1\right\rangle $, with the different
cavity decay rate $\kappa$: $\kappa=\{\gamma_{ea},0.7\gamma_{ea}%
,0.5\gamma_{ea},0.3\gamma_{ea}\}$. (b) the area $S={\displaystyle\int
\nolimits_{0}^{T}}Im[X(t)]dt$ ($T=50/\gamma_{ea}$) in (a). The parameter
$G=0.03\gamma_{ea}$ and other parameters are the same as in Fig. 2.}%
\label{4}%
\end{figure}

\begin{figure}[ptb]
\includegraphics[width=5.0in]{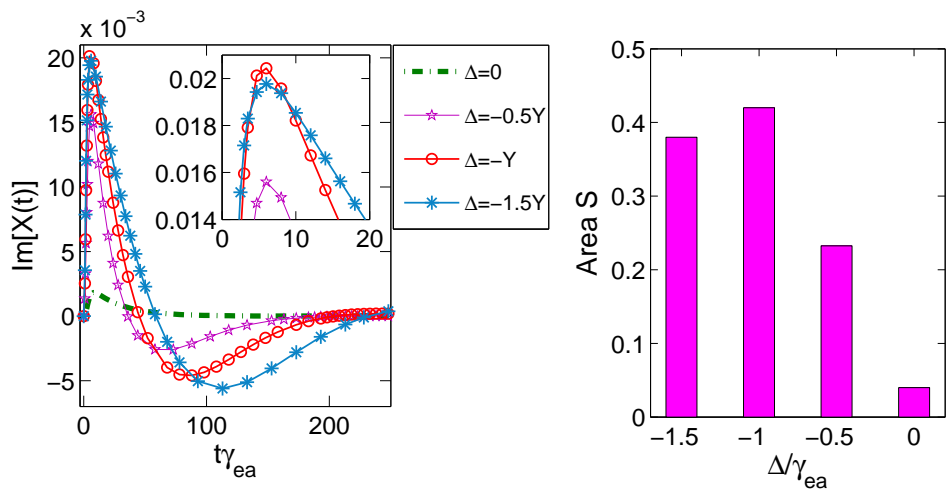}\newline\caption{(Color online)(a) The
imaginary part of time-dependent linear susceptibility $Im[X(t)]$ for the
single-photon Fock state $\left\vert 1\right\rangle $, with the single-photon
detunings $\Delta_{p}=\Delta_{c}=\Delta$: $\Delta=\{0,-0.5\gamma_{ea}%
,-\gamma_{ea},-1.5\gamma_{ea}\}$. (b) the area $S={\displaystyle\int
\nolimits_{0}^{T}}Im[X(t)]dt$ ($T=50/\gamma_{ea}$) in (a). The parameter
$G=0.03\gamma_{ea}$ and other parameters are the same as in Fig. 2.}%
\label{5}%
\end{figure}%
\begin{align}
\frac{d\rho_{a-p}(t)}{dt} &  =-i[H_{1}+H_{2},\rho_{a-p}(t)]\nonumber\\
&  +\frac{\gamma_{ea}}{2}[2\sigma_{ae}\rho_{a-p}(t)\sigma_{ea}-\sigma_{ee}%
\rho_{a-p}(t)-\rho_{a-p}(t)\sigma_{ee}]\nonumber\\
&  +\frac{\gamma_{eb}}{2}[2\sigma_{be}\rho_{a-p}(t)\sigma_{eb}-\sigma_{ee}%
\rho_{a-p}(t)-\rho_{a-p}(t)\sigma_{ee}]\nonumber\\
&  +\frac{\gamma_{deph}}{2}[2\sigma_{ee}\rho_{a-p}(t)\sigma_{ee}-\sigma
_{ee}\rho_{a-p}(t)-\rho_{a-p}(t)\sigma_{ee}]\nonumber\\
&  +\frac{\kappa}{2}[2a\rho_{a-p}(t)a^{\dagger}-a^{\dagger}a\rho_{a-p}%
(t)-\rho_{a-p}(t)a^{\dagger}a],
\end{align}
where $\gamma_{ea}$ ($\gamma_{eb}$) is the spontaneous emission rate from
state $\left\vert e\right\rangle $ to $\left\vert a\right\rangle $
($\left\vert b\right\rangle $), $\gamma_{deph}$ describes energy-conserving
dephasing processes rates, and $\kappa$ is the cavity decay rate. Equation (2)
cannot be solved exactly. We numerically calculate the time-dependent linear susceptibility%

\begin{equation}
X(t)=\frac{N\left\vert \mu_{ae}\right\vert ^{2}\rho_{ea}}{\epsilon_{0}%
\hbar\Omega_{p}},
\end{equation}
with the initial state $\rho_{a-p}(0)=\rho_{a}^{ste}\otimes\left\vert
n\right\rangle $, where $N$ is the atomic density, $\mu_{ea}$ is the dipole
matrix element of the probe transition, $\rho_{ea}$ denotes the off-diagonal
density-matrix element, and $\rho_{a}^{ste}$ is the atomic density operator
under steady-state EIT. We give our attention to the imaginary part of linear
susceptibility $Im[X(t)]$, which characterizes the absorption of atomic ensemble.

Figure 2 shows $Im[X(t)]$ versus the time $t$ in units of $1/\gamma_{ea}$ for
Fock state $\left\vert n\right\rangle $: $n=\{0,1,2,3,4\}$ with different
dispersive coupling strength. From Fig.2, we see that the steady state EIT
changes to the transient dynamics when the photons are injected into the
cavity; furthermore, the peaks of $Im[X(t)]$ changes with the photon number
$n$. Thus one could measure the peak transmission loss of the probe field to
measure the photon number $n$. We can see the origin of this result, via the
stochastic wave-function description of the dynamics of the system with
non-Hermitian effective Hamiltonian \cite{Fleischhauer} $H_{non}=-(\Omega
_{p}\sigma_{ea}+\Omega_{c}\sigma_{eb}+H.c.)+a^{\dagger}a(G\sigma_{bb}%
-i\kappa/2)+(\Delta_{p}-i\gamma_{ea}/2)\sigma_{ee}+\delta\sigma_{bb}$. Under
the two-photon resonant conditions $\delta=0$, the atomic system with the
application of two classical fields moves into the dark state $\left\vert
dark\right\rangle \propto\Omega_{c}\left\vert a\right\rangle -\Omega
_{p}\left\vert b\right\rangle $, which has no population in excited state
$\left\vert e\right\rangle $. Hence there are no spontaneous emission and no
transmission loss of a probe field in the ideal case. Then there is a photon
Fock state $\left\vert n\right\rangle $ ($n\neq0$), which shifts the level
$\left\vert b\right\rangle $ and moves the atomic system out of the dark state
$\left\vert dark\right\rangle $. Since the level shift is proportional to
photon number $n$, the dynamics of EIT departing from dark state $\left\vert
dark\right\rangle $ is explicitly dependent on the photon number $n$.

We can enhance the signal-to-noise ratio for the measurement process involving
accumulated measurement. In Fig.3, we plot the area%

\begin{equation}
S=%
{\displaystyle\int\nolimits_{0}^{T}}
Im[X(t)]dt,
\end{equation}
with the different dispersive coupling strength $G$: $G=\{0.08\gamma
_{ea},0.03\gamma_{ea},0.008\gamma_{ea},0.003\gamma_{ea}\}$ when $T=50/\gamma
_{ea}$. We find that the area $S$ strongly depends on the photon number $n$
even if $G<\kappa,\gamma_{ea}$. The photons are dispersively coupled to the
atoms, and when the photon number is small, $g\sim10G$. Thus the cooperativity
parameter $\eta=g^{2}/\kappa\gamma_{ea}$ of the optical cavity in Fig. 3 (a),
(b), (c), and (d) are $\eta_{a}\sim2$, $\eta_{b}\sim0.3$, $\eta_{c}\sim0.02$,
and $\eta_{d}\sim0.003$, respectively. Figure 3 indicates there are evident
changes of transmission of a probe field with different Fock state $\left\vert
n\right\rangle $, even if the cooperativity parameter $\eta<1$, which means
that a measurement on the changes of transmission of the probe field can be
used for effective QND measurement of the small photon number, even if the
optical cavity is not in strong coupling regime.

Next we explore the influences of cavity decay rate $\kappa$ and the
single-photon detuning $\Delta_{p}=\Delta_{c}=\Delta$ for the single-photon
Fock state $\left\vert 1\right\rangle $. In Fig.4, we investigate the curve
$Im[X(t)]$ and the area $S$ , with the different cavity decay rate $\kappa$:
$\kappa=\{\gamma_{ea},0.7\gamma_{ea},0.5\gamma_{ea},0.3\gamma_{ea}\}$. Figure
4 indicates that there is a visible change of transmission of a probe field,
even when $\kappa=\gamma_{ea}$ and the injected photon is in single-photon
Fock state $\left\vert 1\right\rangle $. Figure 5 shows the curve $Im[X(t)]$
and the area $S$ for the single-photon Fock state $\left\vert 1\right\rangle
$, with the single-photon detunings $\Delta_{p}=\Delta_{c}=\Delta$:
$\Delta=\{0,-0.5\gamma_{ea},-\gamma_{ea},-1.5\gamma_{ea}\}$. From Fig.5, we
see that an appropriate single-photon detuning, e.g. $\Delta=-\gamma_{ea}$,
may effectively increase the total changes of the transmission of the probe field.

We note that Reference \cite{Suzuki} has shown that the average cavity photon
number $\left\langle n_{c}\right\rangle $ can be determined by a measurement
on the peak transparency of the probe field. However, our protocol is much
different from that in Ref. \cite{Suzuki}, in which the average cavity photon
number $\left\langle n_{c}\right\rangle $ is measured and quantum state of
cavity field changes during the measurement even if it is initially in a
certain Fock state.

\emph{Heralded photon Fock states source with QND measurement.---}In the
following part, we consider the pulsed excitation of the cavity. The\emph{
}interaction Hamiltonian for the cavity mode $a$ driven by a laser field is
\cite{Imamoglu1} $H_{dri}=\sqrt{2\kappa}(\beta^{\ast}a-\beta a^{\dagger})$,
where $\beta$ is the laser field amplitude. When $\beta\gg\sqrt{\kappa}$, a
weak coherent field $\left\vert \alpha\right\rangle $ can be generated within
a short time, in which the cavity decay can be neglected. Assuming that a weak
coherent field is injected into the cavity, an observable photon number $n$ is
measured by detecting the change of transmission of the probe field, which
makes the coherent field collapse into a certain Fock state. Thus a probable,
but heralded small photon Fock states in particular single-photon state is
created. After the time $t^{^{\prime}}>200/\gamma_{ea}$, the transient
dynamics of EIT returns to steady state EIT, and if another weak coherent
field is injected into the cavity, the system repeats the above process. Thus
many weak coherent fields are sequently injected into the cavity at time
$t^{^{\prime}}>200/\gamma_{ea}$ interval, and detection of the change of
transmission loss of the probe field will realize a heralded single-photon
source, which well suits for actual QKD \cite{Gisin}.

\emph{Discussion and conclusion.---} The transmitted intensity of a probe beam
that has the wavelength $\lambda$ can be obtained from the relation
$I_{p}=I_{0}e^{-\alpha l}$, here $\alpha=2\pi Im[X(t)]/\lambda$ is the
absorption coefficient, $l$ is the length of a cold atomic sample, and $I_{0}$
is incident intensity of the probe beam. Thus the transmission change of a
classical optical field is $P=%
{\displaystyle\int\nolimits_{0}^{T}}
(I_{p}-I_{0})/I_{0}dt=%
{\displaystyle\int\nolimits_{0}^{T}}
(1-e^{-\alpha l})dt\sim%
{\displaystyle\int\nolimits_{0}^{T}}
\alpha ldt=2\pi Sl/\lambda$ (when $\alpha l\ll1$). Because the area $S$
strongly depends on the photon number $n$, QND measurement can be achieved by
the measurement of the transmission change of a classical optical field. The
measurement time $T=50/\gamma_{ea}\gg1/\kappa$, so this means that our scheme
does not require that the measurement time is shorter than the cavity decay
time $1/\kappa$. There is a visible change of transmission loss of a probe
field, even when the injected photon is in single-photon Fock state
$\left\vert 1\right\rangle $. Thus our scheme provides a heralded
single-photon source for actual QKD \cite{Gisin}. In our scheme, the numerical
calculation shows our scheme still works even if the cooperativity parameter
$\eta<1$, and this may largely lower the experimental requirements.

Now we address the experiment feasibility of the proposed scheme. We consider
an ensemble of $^{87}Rb$ atoms trapped in an optical cavity. The states
$\left\vert a\right\rangle $ and $\left\vert b\right\rangle $ correspond to
$\left\vert F=1,m=-1\right\rangle $ and $\left\vert F=1,m=1\right\rangle $ of
$5S_{1/2}$ ground levels respectively, while $\left\vert e\right\rangle $ and
$\left\vert f\right\rangle $ correspond to $\left\vert F=1,m=0\right\rangle $
and $\left\vert F=1,m=1\right\rangle $ of $5P_{3/2}$ excited level,
respectively. The relevant cavity QED parameters in the experiment are
$(g_{0},\kappa,\gamma_{e})/2\pi=(27,4.8,6)MHz$ \cite{Sauer}, which correspond
to the cooperativity parameter $\eta=g_{0}^{2}/\kappa\gamma_{e}=25\gg1$. Under
the experimental conditions, the principle experiment could be realized.

In summary, we have proposed a method for QND measurement and heralded
preparation of Fock states with dynamics of EIT in an optical cavity. In our
scheme, an atomic medium trapped in an optical cavity driven by two
continuous-wave classical fields moves into a dark state. Then a weak coherent
field enters the cavity, shifts the level $\left\vert b\right\rangle $ and
moves the atomic medium out of the dark state. Because the curve of $Im[X(t)]$
explicitly depends on the number $n$ of photons, a measurement on the changes
of transmission of the probe field can be used for effective QND measurement
of the small photon number and preparation of Fock states. Our numerical
calculation shows the scheme still works in an optical cavity without strong
coupling. Thus our method opens an alternate for QND measurement and
preparation of photon Fock states in particular single-photon state in a
heralded way.

\textbf{Acknowledgments: }This work was supported by the National Natural
Sciences Foundation of China (Grants Nos. 60978013 and 11074263), the Key Basic Research
Foundation of Shanghai (Grant No. 09310210Z1), the Shanghai Commission of
Science and Technology (Grant. No. 10530704800), the Shanghai Rising-Star
Program (Grant No. 11QA1407400), and the China Postdoctoral Science Foundation
(Grant No. 2011M500054).

\end{document}